\documentstyle[aps,amstex,prl,epsfig]{revtex}     

\newcommand{\M}{{\bf {M}}}  
\newcommand{\Ha}{{\bf {H}}}     
\newcommand{\He}{H_{ext}}  
\newcommand{\E}{{\bf {E}}}  
\newcommand{\D}{{\bf {D}}}  
\newcommand{\B}{{\bf {B}}}  
\newcommand{\ve}{{\bf {v}}}  
\newcommand{\na}{\nabla}  
\newcommand{\ti}{\times}  
\newcommand{\pa}{\partial}  
\newcommand{\m}{\mu_r}  
\newcommand{\al}{\alpha}  
\newcommand{\psib}{\overline{\psi}}  
  
\newcommand{\veb}{\overline{\bf v}}  
\newcommand{\wb}{\overline{w}}

\newcommand{\zetab}{\overline{\zeta}}

\begin{document}     
     
\draft     
\title{Dissipation in ferrofluids: Mesoscopic versus hydrodynamic theory}     
\author{ Hanns Walter M\"uller$^{(a)}$ and Andreas Engel$^{(b)}$}     
\address{(a)  Max-Planck-Institut f\"ur Polymerforschung,  
 Ackermannweg 10,  D-55128 Mainz and \\(b) Institut f\"ur Theoretische Physik,
Otto-von-Guericke-Universit\"at Magdeburg, PSF 4120, D-39016 Magdeburg}     

\maketitle     
     
\begin{abstract}     
Part of the field dependent dissipation in ferrofluids occurs due to the
rotational motion  
of the ferromagnetic grains relative to the viscous flow of
the carrier fluid. The classical theoretical description due to Shliomis
uses a mesoscopic treatment of the particle motion 
to derive a relaxation equation for the non-equilibrium part of the
magnetization. Complementary, the hydrodynamic approach of
Liu involves only macroscopic quantities and results in dissipative Maxwell 
equations for the magnetic fields in the ferrofluid. Different stress tensors
and constitutive equations lead to deviating theoretical predictions 
in those situations, where the
magnetic relaxation processes cannot be considered instantaneous on the
hydrodynamic time scale. We quantify these differences for two situations of
experimental relevance namely a resting fluid in an oscillating oblique field 
and the damping of parametrically excited surface waves. The 
possibilities of an experimental
differentiation between the two theoretical approaches is discussed.

\end{abstract}     
  
47.35.+i Hydrodynamic waves
\section{Introduction}  
The interplay between hydrodynamic behavior and magnetic field sensitivity
gives rise to a variety of new physical effects and makes ferrofluids 
a fascinating subject of research with many interesting technical
applications \cite{rosensweig85,bashtovoi88,berkovski93}. 
Practically all hydrodynamic properties can
be modified or modulated by an external magnetic field. Although very similar
effects are possible with electrically polarized fluids, the coupling of the
fields to the hydrodynamics is more pronounced in the magnetic case. Since 
 the experimental handling of strong magnetic fields is simpler than
that of  electric ones, ferrofluids are the experimenters choice to analyze
the field affected  hydrodynamic motion.

On the other hand a thorough theoretical understanding of the many aspects of
ferrofluid motion is a rather ambitious task
\cite{shliomis72,liu93}. 
This is in particular true if the motion is such that dissipation effects are
important. The theoretical analysis must then account for the deviations from
and the relaxation towards equilibrium in a consistent and quantitatively
accurate way.  

One of the interesting properties of ferrofluids is their tunable viscosity
which can be controlled by an external magnetic field.  This so called
magneto-viscous effect has first been observed experimentally by Mc Tague
\cite{mctague69}. Qualitatively it can be accounted for \cite{hall69} by
magnetic torques acting upon the suspended ferromagnetic particles: depending
on the relative orientation between magnetic field and   
local vorticity the particle rotation is hindered. The friction at the   
coated particle surfaces generates an extra dissipation and thus leads to an   
enhanced effective viscosity. This additional rotational friction is   
described by a ``rotational viscosity'' $\eta_R$. 
Based on the theory of rotational Brownian motion of non-interacting rigid
dipoles, Shliomis \cite{shliomis72} developed a theoretical description of
this effect. The basic ingredient is a relaxation equation for the
non-equilibrium part of the magnetization resulting from a 
stochastic description of an ensemble of ferromagnetic grains. This 
treatment allows to derive an
analytic expression for the magnetic field dependence of $\eta_R$ 
\cite{shliomis72}. We will
refer to this approach as the {\it mesoscopic} theory since it starts with a
rather detailed characterization of the processes occurring on a 10 nm scale in 
order to determine macroscopic properties like the viscosity.

Quite complementary the same phenomena can be described 
within the framework of non-equilibrium thermodynamics. Appropriate
magnetic field variables must then be introduced into the set of 
relevant thermodynamic
quantities. New dissipative currents show up when the system is driven out of 
equilibrium. They are coupled to the thermodynamic forces by 
additional Onsager coefficients. This program was carried out by Liu and coworkers in a
series of papers \cite{liu93,liu95,henliu} and gave rise to what is called 
the {\it hydrodynamic} Maxwell theory. Making no reference at all to the 
microscopic mechanism of dissipation this approach has the appealing feature
of being very general and applicable to electrically or magnetically
polarizable continuous media. As in every hydrodynamic theory the Onsager
coefficients are free parameters which have to be determined experimentally
or from a microscopic theory. 

The detailed quantitative relation between the two theoretical approaches is
presently not completely clear and there has been some controversy in the
literature over the recent years. In order to contribute to a clarification of 
this issue we investigate in the present paper the consequences of both the
mesoscopic and the hydrodynamic theory for two simple experimental setups
involving ferrofluids. On the one hand we analyze the influence of an
oscillating oblique magnetic field on a ferrofluid at rest, on the other hand
we investigate the damping of parametrically driven surface waves on the
ferrofluid in a constant external magnetic field perpendicular to the
undisturbed surface. Both situation are easily accessible to
experiments. Interestingly the two theoretical approaches give rise to 
{\it different} results for some of the relevant quantities. We finally
discuss whether these differences are pronounced enough to allow an
experimental decision between the theories.

\section{Equations of motion}  
The fluid motion is governed by the hydrodynamic balance equations for mass
and linear momentum. For an incompressible ferrofluid (density $\rho$) with a
velocity field ${\bf v}({\bf r},t)$ these equations read as  
\begin{eqnarray} 
\nabla \cdot {\bf v}&=0  
\label{conteqn}\\  
\rho(\partial_t {\bf v}+{\bf v}\cdot \nabla){\bf v}&= \nabla \cdot \Pi + \rho
{\bf g},   
\label{nseqn} 
\end{eqnarray}  
where $\Pi_{ij}$ is the stress tensor containing reactive and dissipative
contributions  (see later) and ${\bf g}=(0,0,-g)$ is the acceleration due to
gravity. The two theoretical approaches under consideration differ in their
expressions for the stress tensor and the field equations.
It will turn out that they produce different results when the  magnetic  
relaxation time $\tau$ (usually between $10^{-3}$ and $10^{-6}s$)  
cannot be neglected with respect to the time scale, $1/\omega$ say, of the
hydrodynamic motion. In those cases it is argued that  
the local magnetization ${\bf M}({\bf r},t)$ deviates from its 
local equilibrium value ${\bf M}_{eq}({\bf r},t)$ giving rise to an
additional dissipation, which we shall denote here as "magneto-dissipation".   
This phenomenon is responsible for the field dependent  
viscosity \cite{shliomis72,mctague69,hall69}, the magneto-vortical resonance 
\cite{gazeau96,gazeau97}, or the spectacular negative viscosity effect  
\cite{shliomis94,bacri95,zeuner98} detected recently in ferrofluids.  
 
In order to simplify the mathematics we make throughout this paper the
following approximations which do not remove the basic differences between the 
two approaches: 
(i) The magnetic field ${\bf H}$ within the ferrofluid is weak enough to
warrant a linear relationship ${\bf M}_{eq}=\chi {\bf H}$, where the 
susceptibility is supposed to be proportional to the density $\rho$. 
(ii) The local non-equilibrium magnetization    
${\bf M}$ deviates only slightly from ${\bf M}_{eq}$. 
(iii) Magneto-dissipation is treated in the so called low-frequency limit 
characterized by $\tau\omega\ll 1$.
\subsection{Field equations in the mesoscopic theory}  
Following Shliomis \cite{shliomis72} (see also Ref.~\cite{rosensweig85}) 
the stress tensor for ferrofluids appears
in the form   
\begin{equation}  
\Pi^{mes}_{ij}=-(p+\frac{\mu_0}{2}H^2)\delta_{ij}
+H_i B_j + \eta(\nabla_i v_j +  
\nabla_j v_i) + \frac{\mu_0}{2} \varepsilon_{ijk} ({\bf M} \times  
{\bf H})_k.  
\label{pi_shliomis} 
\end{equation}  
In Eq.~(\ref{pi_shliomis}) occurs the pressure field $p({\bf r},t)$,  
the Maxwell stress tensor with the induction ${\bf B}=\mu_0 ({\bf H}+{\bf M})$
and the  usual dissipative contribution proportional to the shear viscosity
$\eta$. Moreover, the last term on the right hand side of
Eq.~(\ref{pi_shliomis}) arises if the magnetization exhibits an   
off-equilibrium component perpendicular to ${\bf H}$. It is related to the   
magneto-dissipation and guarantees the symmetry of the stress tensor if the
local directions of ${\bf H}$ and ${\bf B}$ do not coincide. 
The appearance of ${\bf M}$ in the stress tensor (\ref{pi_shliomis}) requires
an extra constitutive equation for the
off-equilibrium component $\delta {\bf M}={\bf M} - \chi {\bf H}$ of the
magnetization. Under the above simplifications (i-iii) the following relation
 can be established \cite{bashtovoi88}  
\begin{equation}  
\delta{\bf M}= - \tau \chi \left [ \partial_t {\bf H} + ({\bf v} \cdot \nabla)
  {\bf H} -  \frac{\nabla \times\ {\bf v}}{2} \times {\bf H} \right ],  
\label{delta_m} 
\end{equation}  
where $\tau$ denotes a relaxation time. 

The magnetic part of the problem is treated in the framework of the  
magnetostatic approximation   
\begin{eqnarray}  
\nabla \times {\bf H}=0\\   
\nabla \cdot {\bf B}=0. 
\label{magnetostatic}  
\end{eqnarray}  
With this simplification and the above expression for the stress tensor 
(\ref{pi_shliomis})  the hydrodynamics obeys the Navier-Stokes equation, 
which assumes the form  
\begin{equation} 
\rho \partial_t {\bf v}+\rho ({\bf v}\cdot \nabla){\bf v}= 
-\nabla p + \rho {\bf g} + \eta \nabla^2 {\bf v} + \mu_0 ({\bf M} \cdot \nabla) {\bf H} 
+ \frac{\mu_0}{2} \nabla \times ({\bf M}\times {\bf H}). 
\label{ns_shliomis} 
\end{equation} 
\subsection{The field equations of the hydrodynamic theory}  
The complete set of hydrodynamic Maxwell equations is given by \cite{liu93}
\begin{eqnarray}  
\label{maxequ}  
\pa_t \D = \na\ti\Ha-{\bf j}_{el}\qquad &\pa_t \B=-\na \ti\E\\  
\na \cdot \D=\rho_{el}\qquad\qquad\qquad\quad   &\na \cdot \B=0, \nonumber  
\end{eqnarray}  
where ${\bf E}$ and ${\bf D}$ denote the electric and displacement fields. For
non-conducting ferrofluids there are no electric charges or currents, thus   
${\bf j}_{el}=0$ and $\rho_{el}=0$.  
The fields $\Ha$ and $\E$ split into reactive and dissipative parts   
\begin{equation}  
\Ha=\Ha^R+\Ha^D \qquad\text{and}\qquad \E=\E^R+\E^D\nonumber  
\end{equation}  
where $\Ha^R=\pa \epsilon/\pa \B$ and $\E^R=\pa \epsilon/\pa \D$ are    
derivatives of the thermodynamic energy density $\epsilon$ and    
contain only equilibrium information. The dissipative fields $\Ha^D$ and   
$\E^D$ represent the off-equilibrium currents and are functions of the   
thermodynamic forces.   
For isotropic isothermal situations and neglecting off-diagonal   
couplings they depend only on the electromagnetic fields in the local rest   
frame and are given by \cite{liu93}  
\begin{equation}\label{maxirre}  
\Ha^D=-\frac{\alpha}{\mu_0}\na\ti(\E^R+\ve\ti\B)\qquad \text{and} \qquad  
\E^D=\frac{\beta}{\epsilon_0}\na\ti(\Ha^R-\ve\ti\D),  
\end{equation}   
where the parameters $\alpha$ and $\beta$ are Onsager coefficients.   
In the present paper we restrict ourselves to ferrofluids with a negligible
dielectric susceptibility $\chi_{el}\simeq0$ (e.g. hydrocarbon based
ferrofluids) giving rise to $\beta\simeq 0$ (see also \cite{liu98}) implying
$\E^D=0$. Moreover, the ratio between time and space
derivatives of fields respectively is $\omega/(c k)$ and for hydrodynamic
frequencies $\omega\lesssim 10^3 s^{-1}$ and wavelengths 
$1/k \sim 10^{-3}m^{-1}$ 
we may neglect the time derivative of $\D$ in the Maxwell equations
(\ref{maxequ}). This means that the magnetic field may be assumed to follow
the fluid motion  instantaneously. In this way we recover 
the magnetostatic field equations   
\begin{eqnarray}  
\label{Hequ1}  
\na \cdot \B&=0\\  
\na\times \Ha&=0 \nonumber\quad.  
\end{eqnarray}  
At a liquid-air interface these fields are subjected to the usual continuity
conditions for the normal component of $\B$ and the tangential component of
${\bf H}$.    
  
Using incompressibility we get from (\ref{maxequ}) and (\ref{maxirre})  the
following constitutive relation for the dissipative field   
\begin{equation} 
\label{HDequ}  
\Ha^D=\frac{\al}{\mu_0}\left [ \pa_t\B + (\ve \cdot \na)\B - (\B \cdot \na)\ve
\right ] \quad,   
\end{equation}  
which is the analogue to $\delta {\bf M}$ in the approach of Shliomis. For
later use we introduce the {\em reactive} contribution to the magnetization   
\begin{equation}  
\label{defM}  
\M^R=\frac{1}{\mu_0}\B-\Ha^R \quad.  
\end{equation}  
This definition differs from the ``real'' magnetization $\M=\B/\mu_0-\Ha$   
in using only the reactive part $\Ha^R$ of the magnetic field. The   
advantage of this formulation is that for the present case of constant and
scalar susceptibility the fields $\B, \Ha^R$ and $\M^R=\chi {\bf H}^R$ are all
parallel to each other \cite{footnote1}. 
The fact that $\Ha$ and hence $\M$ are not parallel to
$\B$ is the very reason for the magneto-dissipative effects    
\cite{shliomis72,mctague69,liu93} investigated here.  
  
With the assumption of the susceptibility being proportional to the density
$\rho$, the stress tensor of the fluid simplifies to \cite{liu93,LL}  
\begin{equation}  
\label{pi_liu}  
\Pi^{hyd}_{ij}=-\left [ p + \frac{\mu_0}{2}(H^R)^2 \right ] \delta_{ij} 
+ H_i^R B_j +   
   \eta(\nabla_j v_i+\nabla_i v_j ) \quad.  
\end{equation}  
The fluid dynamics is therefore governed by the Navier-Stokes equation, which
using (\ref{pi_liu}) acquires the form  
\begin{equation}  
\label{NSE}  
\rho \partial_t \ve + \rho(\ve\na)\ve =   
  -\na p +\rho {\bf g} + \eta \nabla^2 \ve + \mu_0 M^R \na H^R +
  \B\ti(\na\ti\Ha^D).      
\end{equation}  
The fourth term on the r.h.s. of the Navier-Stokes equation was simplified to
involve only the magnitudes of the vectors $\M^R$ and $\Ha^R$ by exploiting
the fact that they are parallel.  
\subsection{Comparison between the two approaches} 
\label{comparison}  
Let us first emphasize that the two descriptions introduced above are fully
equivalent if the magnetic relaxation processes can be considered to be 
instantaneous ($\tau=\alpha=0$) on the time scale of the hydrodynamic motion.  
In this case of vanishing magneto-dissipation we have 
$\delta {\bf M}={\bf H}^D=0$, i.e. ${\bf M}^R={\bf M}\parallel {\bf H}={\bf H}^R$, and
thus the stress tensors (\ref{pi_shliomis}) and (\ref{pi_liu}) or respectively
the Navier-Stokes equations (\ref{ns_shliomis}) and (\ref{NSE}) coincide.  
 
Differences between the two approaches arise due to the treatment of the  
magneto-dissipation occurring at finite  $\delta {\bf M}$ and ${\bf H}^D$,
respectively. By comparing  Eqs.~(\ref{pi_shliomis}) and (\ref{pi_liu})  
and expressing the antisymmetric part of the tensor element $H_i^D B_j$ in
(\ref{pi_liu}) by 
$(1/2)\varepsilon_{ijk}({\bf H}^D \times {\bf B})_k= 
-(\mu_0/2)\varepsilon_{ijk}({\bf M}\times  
{\bf H})_k$ one arrives at  
\begin{equation}  
\Pi^{hyd}_{ij} - \Pi^{mes}_{ij}=\mu_0 {\bf H}\cdot {\bf H}^D \delta_{ij}  
-\frac{1}{2} (H_i^D B_j + H_j^D B_i), 
\label{pi_diff}  
\end{equation}  
where terms of order $(H^D)^2$ have been dropped.
The  stress tensors in Eq.~(\ref{pi_diff}) differ by a diagonal element and  a 
symmetric non-diagonal contribution. The isotropic term just leads to a
re-normalization of the pressure, which we won't pursue here. The non-diagonal
tensor, however, gives rise to a dissipative force which might be detectable
in an experiment (see below).   
  
It is also instructive to compare the constitutive equations, which determine
the dissipative fields: From the identity  
\begin{equation}  
\mu_0 ({\bf H}+{\bf M})={\bf B}=\mu_0 ({\bf H}^R + {\bf M}^R)=\mu_0 (1+\chi)
{\bf H}^R   
\label{identity} 
\end{equation}  
one finds that the dissipative field contributions $\delta{\bf M}$ and 
${\bf H}^D$ coincide up to a pre-factor:
\begin{equation}  
{\bf H}^D=-\frac{1}{1+\chi} ({\bf M}- \chi {\bf H}) = 
 -\frac{1}{1+\chi} \delta {\bf M}. 
\label{relation_HD_dM}  
\end{equation}  
The associated constitutive relations   
\begin{align}  
\delta{\bf M}&=-\tau \chi \left [ \partial_t {\bf H} + ({\bf v} \cdot \nabla)
  {\bf H}- \frac{1}{2}(\nabla \times {\bf v}) \times  {\bf H} \right ] 
\label{konstit_s}\\  
{\bf H}^D&=\frac{\alpha}{\mu_0}  [ \partial_t {\bf B} + ({\bf v} \cdot
\nabla){\bf B} - ({\bf B} \cdot \nabla) {\bf v}  ], 
\label{konstit_l}  
\end{align}  
however, are not fully compatible: Omitting contributions of
O$(\alpha\omega)^2$ Eq.~(\ref{konstit_l}) can be re-casted as 
\begin{equation} 
{\bf H}^D=\alpha (1+\chi)  [ \partial_t {\bf H} + ({\bf v} \cdot \nabla){\bf H} -  
({\bf H} \cdot \nabla) {\bf v}  ]. 
\label{konstit_l2} 
\end{equation} 
By comparing the two first terms on the right hand side of
Eqs.(\ref{konstit_s}, \ref{konstit_l2}) and using (\ref{relation_HD_dM}) 
one can relate the characteristic relaxation times by \cite{footnote2} 
\begin{equation}  
\alpha= \tau \frac{\chi}{(1+\chi)^2}.  
\label{times} 
\end{equation}  
The third terms in Eqs.~(\ref{konstit_s},\ref{konstit_l2})   
describe different physics. This is easy to see by considering a homogeneous
stationary magnetic field: According to the mesoscopic theory, 
Eq.~(\ref{konstit_s}), only a rotational flow is able to drive a
magneto-dissipative field. Contrary in the hydrodynamic approach,
Eq.~(\ref{konstit_l2}) states that a flow gradient parallel to the magnetic
field suffices to excite a finite ${\bf H}^D$. 
It is interesting to point out that the deviation between the constitutive equations 
Eqs.(\ref{konstit_s}, \ref{konstit_l2}) drops out for a solid body rotation
flow field ${\bf v}=\omega \times {\bf r}$, which is often investigated in the
ferrofluid literature.   
  
To summarize this section: We observe two differences between the classical 
description of Shliomis and the hydrodynamic Maxwell theory of Liu: One in the 
stress tensors $\Pi_{ij}$ and a second in the constitutive equations 
for the dissipative fields $\delta {\bf M}$ and ${\bf H}^D$ respectively. 
In the following we investigate the possibilities of an experimental
differentiation between the two theoretical approaches.  
\section{Ferrofluid at rest in an oblique oscillating magnetic field} 
 We consider a resting ferrofluid layer exposed to a linearly 
 polarized magnetic field ${\bf H}^{(2)}=(H_x^{(2)},0,H_z^{(2)})\cos{\omega t}$  
directed oblique to the surface.   
Here and in the following fields in the air carry the superscript  $"(2)"$,  
while fields within the ferrofluid are denoted without superscript.

The stress tensor in the air above the ferrofluid is taken as
\begin{equation}  
\Pi^{(2)}_{ij}=-\left [ p^{atm} + \frac{\mu_0}{2}(H^{(2)})^2 \right ]
\delta_{ij} + H^{(2)}_i B^{(2)}_j, 
\label{air_stress}
\end{equation} 
where we assume a constant atmospheric pressure $p^{atm}$. 
 
For a fluid at rest the constitutive equations 
(\ref{konstit_s},\ref{konstit_l})   
coincide and thus eventual discrepancies between the two theories will result from
the deviating stress tensors. To test possible differences we   
evaluate the tangential force $f_t=\Pi_{xz}-\Pi_{xz}^{(2)}$ acting 
upon a free surface element. By using the  
magnetic interface conditions for ${\bf H}$ and ${\bf B}$ and owing to the  
fact that isotropic tensor elements do not contribute to tangential forces we
obtain  
\begin{mathletters} 
\begin{equation} 
f_t^{mes}=\frac{\mu_0}{2}(\delta M_xH_z-\delta M_zH_x)= 
-\frac{\mu_0}{2} \tau \chi \left [ H_z \partial_t H_x - H_x \partial_tH_z
\right ]  \label{ft_s}
\end{equation} 
and 
\begin{equation} 
f_t^{hyd}=-H^D_xB_z=-\frac{\alpha}{\mu_0}B_z \partial_t B_x. 
\label{ft_l} 
\end{equation} 
\end{mathletters} 
The computation of $f_t$ requires knowledge of ${\bf H}$ and  
${\bf B}$ within the ferrofluid. From the respective constitutive equations
and the interface conditions we find 
\begin{mathletters} 
\begin{align} 
{\bf H}^{mes}&=\frac{1}{2}\left [ H_x^{(2)},0,\frac{H_z^{(2)}}{1+\chi-i\omega \tau
    \chi} \right ] \,  e^{i \omega t} + c.c.   \label{h_feld}\\ 
{\bf B}^{hyd}&=\frac{\mu_0}{2} \left [  
\frac{H_x^{(2)}}{\frac{1}{1+\chi}+i \omega \alpha},0,H_z^{(2)} \right ] 
e^{i\omega t} + c.c.\,. \label{b_feld} 
\end{align} 
\end{mathletters} 
Evaluating the tangential stresses (\ref{ft_s},\ref{ft_l}) up to leading order 
in $\tau \omega$ or respectively $\alpha \omega$ we get  
\begin{mathletters} 
\begin{equation} 
 f_t^{mes}=O[(\tau \omega)^2] \label{ft_s-result}
\end{equation} 
and 
\begin{equation} 
 f_t^{hyd}=\frac{\mu_0}{2}\, \alpha \omega \, (1+\chi) \,
        H_x^{(2)}H_z^{(2)}\, \sin{2 \omega t} + O[(\alpha \omega)^2].  
\label{ft_l_result} 
\end{equation} 
\end{mathletters} 
At the considered accuracy level, the mesoscopic approach states that the
resting ferrofluid is in equilibrium, while the hydrodynamic Maxwell theory 
predicts a residual tangential surface force oscillating with 
twice the excitation frequency.  
This force drops out in the static limit $\omega \to 0$ or  
when the applied field is directed  normally  or tangentially to the
surface. The maximum effect is achieved for an inflection angle of $45^\circ$,
where the product  of the incident field components $H_x^{(2)}H_z^{(2)}$ is 
maximum. We conclude that the resting fluid cannot be in equilibrium,  
i.e. the force will drive a convective motion. With the aid of
appropriate tracer particles exposed to the surface of the fluid we
expect the acceleration to be easily detectable in an experiment.  
\section{Surface waves with magneto-dissipation}  
In this section we investigate a setup, which aims at probing the different  
constitutive relations (\ref{konstit_s},\ref{konstit_l}).  
As outlined in Sec.~\ref{comparison} a non-rotational flow profile is most
appropriate for this purpose. Surface waves on a deep fluid are a classical
example: The associated flow field is purely potential,  
except in a thin boundary  layer along the system boundaries.  
By increasing the system dimensions this parasitic influence can be limited to 
the viscous skin layer beneath the free surface. Our aim is to derive the 
complex dispersion relation for low frequency small amplitude surface waves 
with magneto-dissipation
taken into account. 
\subsection{Surface waves without magneto-dissipation}  
In this subsection we review the present knowledge on small amplitude surface 
waves on a ferrofluid in a static magnetic field ${\bf H}_{ext}$ 
normal to the surface. With $z=\zeta(x,t)\propto e^{i(kx-\omega t)}$ we denote
the surface deflection in vertical z-direction depending on the horizontal
coordinate $x$ and on time $t$. The real part of the complex valued function
$\omega(k)=\omega'(k)+ i \omega''(k)$ reflects the wave dispersion, while   
the imaginary part accounts for the decay rate.  
 
For a non-viscous ($\eta=0$)  ferrofluid $\omega''(k)$  vanishes  identically  
and the inviscid wave dispersion (indicated by the subscript $"0"$)  is given
by \cite{cowley67,rosensweig85}  
\begin{equation}  
\omega_0^2(k,H_{ext})=
g k - \frac{\mu_0}{\rho}\frac{\chi^2}{(1+\chi)(2+\chi)}\,H_{ext}^2 \, k^2 +  
\frac{\gamma}{\rho} k^3 . 
\label{inviscid_dispersion} 
\end{equation}  
Here $\gamma$ denotes the coefficient of surface tension.  
The magnetic field leads to a negative contribution proportional to $k^2$. If
$H_{ext}$ is sufficiently strong $\omega_0^2$ becomes negative,   
indicating the onset of the Rosensweig instability.   
 
The effect of viscosity on the dispersion has been investigated  
theoretically in \cite{weilepp96,abou97} under the assumption of an infinitely
fast magnetic relaxation (vanishing magneto-dissipative  effect).  
It is found that the expression for $\omega(k)$ can   
no longer be given explicitly, rather the following implicit relation applies, 
\begin{equation}  
{\cal D}(k,\omega)=\omega^2 - \omega_0^2(k) + X(k,\omega) =0  
\label{implicit_dispersion} 
\end{equation}  
where the viscous contribution  reads  
\begin{equation}  
X(k,\omega)=4 i \omega \nu k^2 + 4 (\nu k^2)^2\{\frac{q}{k}-1\}.  
\label{damping_dispersion} 
\end{equation}  
Here $\nu=\eta/\rho$ is the kinematic viscosity and 
$q=\sqrt{k^2-i \omega/\nu}$.   
Note that the viscous contribution does not  depend on the magnetic
field and therefore coincides with the expression for non-magnetic fluids
\cite{kumar94}. This is because the magnetic contributions to the 
stress tensor are
purely conservative at vanishing $\delta{\bf M}$ or ${\bf H}^D$. In the
mathematical derivation of (\ref{implicit_dispersion},\ref{damping_dispersion})
 this is reflected by a
decoupling of the magnetostatic from the hydrodynamic problem. Later we will
see, that this simplification no longer applies if magneto-dissipative
corrections are taken into account.   
  
\subsection{Formulation of the problem with magneto-dissipation included}  
The investigation requires solving the equations of motion together with
appropriate boundary conditions at the free liquid air interface.   
We consider a horizontally unbounded ferrofluid layer of infinite depth   
(from $z=0$ to $z=-\infty$) in contact with air at its free surface. The layer
is exposed to an external  stationary magnetic field 
${\bf H}_{ext}=(0,0,H_{ext})$ directed  perpendicularly to the undisturbed
surface [see Fig.~1].   
 
The balances of the normal and tangential stresses at the liquid air  
interface are expressed by the conditions   
\begin{align}  
n_i(\Pi_{ij}-\Pi^{(2)}_{ij})n_j&= -\gamma (\nabla_j n_j) n_i 
\label{n_stress} \\ 
t_i(\Pi_{ij}-\Pi^{(2)}_{ij})n_j&= 0 
\label{t_stress}  
\end{align}  
where ${\bf t}({\bf r},t)$ and ${\bf n}({\bf r},t)$ denote local  
unit vectors tangential and normal to  the surface respectively. The 
stress tensor in the air above the ferrofluid is given by 
Eq.(\ref{air_stress}). The discontinuity of the normal
stress condition arises by virtue of the finite surface tension $\gamma$.   
The kinematic surface condition requires  
\begin{equation}  
\partial_t \zeta + u \partial_x \zeta= w,  
\label{kinemat} 
\end{equation}  
where ${\bf v}=(u,0,w)$.  The second lateral space direction $y$ need not to
be considered. 
  
As usual the boundary conditions for the magnetic field  at the free surface
read as   
\begin{eqnarray}  
({\bf B}-{\bf B}^{(2)})\cdot {\bf n}&=0,  
\label{b_cont}\\  
({\bf H}-{\bf H}^{(2)})\cdot {\bf t}&=0. 
\label{h_cont}   
\end{eqnarray}  
\subsection{The motionless state and linearized equations for the
  perturbations}   
As reference state (index $0$) we consider the motionless 
(${\bf v}=0$) fluid
with a flat ($\zeta=0$) surface. The associated magnetic quantities are 
\begin{align}  
\label{basic}  
\B_0=\B^{(2)}_0&=(0,0,\mu_0 \He)\\\nonumber  
\M_0&=(0,0,\frac{\chi}{1+\chi}\He)\\\nonumber 
\Ha_0&=(0,0,\frac{1}{1+\chi}\He)\\\nonumber  
\Ha^{(2)}_0&=(0,0,\He)\\  
\Ha^D&=\delta {\bf M}=0.\nonumber  
\end{align}  
>From the momentum balance with the appropriate boundary condition  we get for
the pressure   
\begin{equation}  
p_0(x,z)=-\rho g z -\frac{\mu_0}{2} \left [ \frac{\chi}{1+\chi} \right ] ^2 \,
\He^2 .   
\end{equation}  
Next we study the time evolution of small perturbations by  
linearizing the field equations   
and their boundary conditions around the  basic state solution. To this end we 
introduce small deviations denoted by $\ve=(u,0,w)$, $\delta p$, $\zeta$,   
${\bf b}$, ${\bf b}^{(2)}$, ${\bf h}$, ${\bf h}^{(2)}$, ${\bf m}$.   
Linearizing the Navier-Stokes equations (\ref{ns_shliomis}) and 
(\ref{NSE}) we respectively find for the mesoscopic and the hydrodynamic
approach 
\begin{equation}  
\rho\pa_t\ve =-\na \delta p +\eta\Delta \ve +\mu_0 M_0 \partial_z  {\bf h} +
\frac{\mu_0}{2} \nabla \times \left [ {\bf M}_0 \times {\bf h} +   
{\bf m} \times {\bf H}_0 \right ] 
\label{ns_s} 
\end{equation}
and 
\begin{equation}
\rho\pa_t\ve =-\na \delta p +\eta\Delta \ve  
  +\mu_0 M_0\na h_z + \mu_0\He (\na H^D_z-\pa_z\Ha^D). 
\label{ns_l} 
\end{equation}  
Here we have exploited the fact that $\Ha_0$ and $\M_0$ are both
parallel to the $z$-direction. The linear magnetostatic field equations
remain unchanged   
\begin{eqnarray}  
0&=\na \cdot  {\bf b}=\na \cdot {\bf b}^{(2)} 
\label{b_bulk}\\  
0&=\na\times {\bf h}=\na\times {\bf h}^{(2)}. 
\label{h_bulk}  
\end{eqnarray}  
The boundary conditions (\ref{b_cont},\ref{h_cont}) 
are linearized by using the lowest order expressions   
\begin{equation}  
{\bf n}=(-\frac{\pa\zeta}{\pa x},0,1)\qquad\text{and}\qquad   
           {\bf t}=(1,0,\frac{\pa\zeta}{\pa x})  
\end{equation}  
to get   
\begin{equation}  
\label{b_bound}  
b_z=b_z^{(2)},  
\end{equation}  
\begin{equation} 
\label{h_bound}  
h_x-h^{(2)}_x=M_0\frac{\pa \zeta}{\pa x}  
\end{equation}  
for the magnetic interface conditions. The kinematic surface condition
simplifies to    
\begin{equation} 
\label{kinemat_lin} 
\partial_t \zeta = w, 
\end{equation} 
while we get  
\begin{equation} 
0=-\rho g\zeta + \delta p +\mu_0 M_0 m_z-2\eta\pa_z w
   +\gamma \partial_x^2 \zeta 
\label{normal_lin}  
\end{equation} 
for the normal stress balance at the surface. 
Introducing the dimensionless magneto-dissipative perturbation parameter 
\begin{equation} 
\kappa= \frac{\alpha \mu_0}{\eta}H_{ext}^2=
\frac{\tau \mu_0 \chi}{\eta (1+\chi)^2}
H_{ext}^2.  
\label{perturb} 
\end{equation} 
the condition for the tangential stress reads in the mesoscopic 
approach 
\begin{equation}  
\frac{\kappa}{4} \, (\partial_x w-\partial_z u) = (\pa_z u + \pa_x w)  
\label{tangent_s_lin}
\end{equation}
whereas for the hydrodynamic Maxwell theory we simply find 
\begin{equation}  
B_0 H^D_x = \eta(\pa_z u + \pa_x w).
\label{tangent_l_lin} 
\end{equation} 

To elucidate the time evolution of the perturbed state we assume for all
lateral field dependencies a plane wave behavior 
$\propto \exp[i(kx-\omega t)]$,  e.g., $\ve=\veb(z) \exp[i(kx-\omega t)]$
or $\zeta={\bar \zeta} \exp[i(kx-\omega t)]$, etc.  
The associated z-dependence (if appropriate) is indicated by a bar. 
%
 %
We now first determine the magnetic fields from (\ref{b_bulk},\ref{h_bulk})
in terms of the surface deflection $\zeta$ and the flow field ${\bf v}$ using
the boundary conditions (\ref{b_bound},\ref{h_bound}). Then  
we solve for the velocity field and impose the hydrodynamic boundary   
conditions (\ref{tangent_s_lin}) and (\ref{tangent_l_lin}), respectively.  
The condition for non-trivial solutions of (\ref{normal_lin}) finally    
yields the desired complex dispersion relation $\omega(k)$. The explicit
calculations are sketched in the appendix.

\subsection{Results} 
\label{results} 
The expressions we find for the implicit dispersion relation ${\cal D}(k,\omega)$ are
of the form 
\begin{align} 
0={\cal D}^{mes}(\omega,k)&=  
\omega^2 - \omega_0^2(k) + 4 i \omega \nu k^2  + 4 (\nu k^2)^2 \{\frac{{\tilde
    q}}{k}-1 \}  - \kappa (\nu k^2)^2 
     \frac{\chi}{2+\chi}  \{ \frac{{\tilde q}}{k}-1 \}   
\label{D_s}\\ 
0={\cal D}^{hyd}(\omega,k)&= 
(\omega^2 +2i\omega\nu k^2)\frac{1-\frac{2\nu k^2}{i\omega}
  +\kappa\frac{\nu k^2}{i\omega}\frac{\chi}{\chi+2}}{1+\kappa\frac{\nu 
    k^2}{i\omega}} -\omega_0^2(k)+
2(2+\kappa)\nu^2 k^3 \hat{q}\frac{1+\frac{\kappa}{\chi+2}}{1+\kappa\frac{\nu 
    k^2}{i\omega}}
\label{D_l} 
\end{align} 
with 
\begin{align}
{\tilde q}^2&=k^2-\frac{i \omega}{\nu (1+\kappa/4)}\\  
{\hat   q}^2&=\frac{k^2-\frac{i\omega}{\nu}}{1+\kappa}. 
\end{align}
We analyze these results in the following for a ferrofluid with a magnetic
susceptibility and viscosity appropriate for surface wave experiments
(e.g. APG J12 
\cite{footnote3}, Ferrofluidics). In a moderate external field up to
$H_{ext}=15 kA/m$, which is below the Rosensweig threshold,  
the constant susceptibility approximation holds with an error better than
$4\%$. At this field strength the  perturbation  parameter given by  
Eq.~(\ref{perturb})  is smaller than $0.06$. 
For a quantitative investigation of the magneto-dissipative effect  
 it is therefore sufficient to expand ${\cal D}(k,\omega)$ up to  
first order in $\kappa$  giving 
\begin{equation} 
{\cal D}(k,\omega)=\omega^2 - \omega_0^2 + X(k,\omega) + \kappa \,
X_M(k,\omega,H_{ext})+ O(\kappa^2)
\end{equation} 
with the magneto-dissipative contributions
\begin{equation} 
X^{mes}_M(k,\omega,H_{ext})=  (\nu k^2)^2 \left \{ \frac{i \omega}{2 \nu k q} 
- \frac{\chi}{2+\chi} \{ \frac{q}{k}-1 \} \right \}  
\label{XM_s}
\end{equation}
in the mesoscopic theory and 
\begin{equation} 
X^{hyd}_M(k,\omega,H_{ext})=\frac{2i\omega\nu k^2}{2+\chi} +
2 (\nu k^2)^2(\frac{q}{k}-1)\left [ \frac{4+\chi}{2+\chi} + 
2 i \frac{\nu k^2}{\omega} \right ] 
\label{XM_l} 
\end{equation} 
in the hydrodynamic approach. We observe a different scaling behavior with the 
shear visosity $\nu$. The magneto-dissipative correction according to 
Eq.~(\ref{XM_s}) scales
with $\nu^{3/2}$, while it is only proportional to $\nu$  in
Eq.~(\ref{XM_l}). This deviating proportionality traces back to the 
constitutive equations. According to Eq.~(\ref{konstit_s}) magneto-dissipation
is associated with rotational flow and thus confined to the thin  
boundary layer beneath the surface,  
where damping is proportional to  $\nu^{3/2}$ \cite{muller97}.  
This is in contrast to Eq.~(\ref{konstit_l}), where  
the dissipative field ${\bf H}^D$ is finite over the whole fluid
layer. In particular, dissipation within the convective bulk, which is known \cite{LL6} 
to scale with $\nu$, provides the leading contribution to Eq.~(\ref{XM_l}).   
 
Due to the smallness of $\kappa$ the  magneto-dissipative contribution  
$ \kappa X_M(k,\omega)$ modifies the ordinary viscous shear damping
$X(k,\omega)$  only slightly. Moreover, when studying the field dependence of
the damping rate $\omega''(k,H_{ext})$ one has to account for appreciable
wavenumber shifts resulting from  the non-dissipative magnetic contribution  
\begin{equation} 
-\frac{\mu_0}{\rho} \frac{\chi^2}{(1+\chi)(2+\chi)} \, H_{ext}^2 k^2, 
\label{magnetic_w} 
\end{equation} 
in ${\cal D}(k,\omega)$. Since $X(k,\omega)$ in turn is largely $k$-dependent,  
the reactive term (\ref{magnetic_w}) strongly feeds back into  the effective 
field 
dependence of $\omega''(H_{ext},k)$ and therefore masks the tiny
magneto-viscous contributiony $X_M$. To circumvent this difficulty 
one has to adapt the wavenumber $k$ such 
that $\omega_0^2(k,H_{ext})$  as given by Eq.(\ref{inviscid_dispersion}) is held constant 
when varying  $H_{ext}$. This 
can be accomplished by simultaneously tuning the wave excitation frequency $\omega'$ during
the  $H_{ext}$-scan. 
Fig.~2a depicts the field dependence of 
$\omega''$ evaluated according to this protocol. As expected, 
the damping rate increases with the field strength. The predictions of the
two approaches deviate by less than $4\%$ (Fig.~2a,b). 
With regard to the approximations made and the uncertainty of the 
material specifications this tiny difference is presumably 
too small to be resolved in a  surface wave experiment. 
For completeness we show in Fig.~2c,d how the excitation 
frequency $\omega'$ must 
be adapted to guarantee a constant $\omega_0$ during the variation of the  
field intensity.

\section{Summary and discussion} 
The aim of the present paper was to compare two different theoretical
approaches to the problem of magnetic field dependent dissipation in
ferrofluids. On the one hand we considered the theory given by 
Shliomis resting on a mesoscopic treatment of the rotation of the magnetic
particles. On the other hand we applied the hydrodynamic Maxwell theory
advocated by Liu which treats field dissipation within the standard framework
of irreversible thermodynamics.  
For experimental arrangements in which the magnetic relaxation processes  
can be considered infinitely fast the two descriptions are
equivalent. This is, e.g., the case if the ferrofluid is composed of
ferromagnetically soft particles, where the associated magnetic relaxation
is usually very rapid as it is dominated by the fast N\'eel process. 
 
On the other hand, for ferrofluids composed of  materials
with a high specific magnetic anisotropy (e.g. elementary cobalt or
cobalt-ferrites) the magnetic relaxation process is  
largely dominated by the Brownian rotation of the particles in the carrier 
matrix. In high viscosity carrier fluids the associated Brownian relaxation 
time can be large and reach the time scale of the hydrodynamic motion.
This is the situation in which magneto-dissipation  must be 
accounted for. We have shown that this effect is treated differently by the
approaches of Shliomis and Liu due to the use of different stress tensors as
well as distinct constitutive equations for the dissipative fields. 

In a ferrofluid layer at rest the constitutive equations coincide. If the
surface is exposed to an oblique linearly polarized magnetic field, the
differing stress tensors lead to different statements about whether the resting
ferrofluid is in equilibrium or not: Whereas the mesoscopic approach gives no
deviation from equilibrium the hydrodynamic theory predicts an onset of
convection.  We expect these predictions to be easy to verify or falsify in an
experiment.   

On the other hand, an experimental check of the validity of the constitutive
equations is probably more subtle. The magneto-dissipative
contribution is usually very small as compared with the ordinary shear
dissipation. Clearly, it becomes most pronounced if the ferrofluid is in solid
body rotation, where  shear dissipation drops out. Unfortunately, this
special flow geometry  -- when introduced into the constitutive equations of
Shliomis and Liu -- leads to the {\it same} dissipative fields. We therefore
looked for a non-rotational flow geometry, which is easy to realize and which
fulfills the requirements of small amplitude low frequency motion. As a
possible candidate we propose the flow field of free surface waves. On the
basis of the two 
theoretical approaches at hand we worked out the associated complex dispersion
relations. In the final expressions for the magneto-dissipative contribution
we observe a different scaling behavior with the shear viscosity $\nu$, 
which could be traced back to the deviation between the constitutive equations.  
On the basis of the material specifications for a real ferrofluid and the
experimental parameters for a low frequency surface wave experiment 
we computed the damping rate and evaluated the difference between the
predictions. Unfortunately, the resulting deviation is less than $4\%$ and
therefore  smaller than the expected experimental error.
\\[.2cm]
{\bf Acknowledgments:} We have largely benefited from stimulating
discussions with Mario Liu. This work is supported by the Deutsche
Forschungsgemeinschaft under SFB 277 and project En 278/2. A.~E. would like to 
thank the {\it Center for Nonlinear Dynamics} at the University of Texas at
Austin where part of the work was done for the kind hospitality and
acknowledges financial support from the {\it Volkswagenstiftung}. 

\section{appendix}
In this appendix we sketch the derivation of the dispersion relations for
surface waves within the two theoretical approaches. 
\subsection{The mesoscopic approach}
Due to (\ref{h_bulk}) we introduce the scalar magnetic potentials by 
\begin{equation} 
\Ha=-\na \psi, \quad 
\Ha^{(2)}=-\na \psi^{(2)}, 
\end{equation} 
which obey the following relations  
\begin{equation} 
\nabla^2 \psi = -\frac{\chi \tau}{2 (1+\chi)} H_0 \nabla^2 w \quad , \quad
\nabla^2 \psi^{(2)}=0 . 
\label{laplace}
\end{equation} 
Eq.(\ref{laplace}) has been derived by using (\ref{magnetostatic}) and 
the constitutive equation (\ref{konstit_s}). 
Invoking once more the low frequency approximation, 
the latter occurs in the form 
\begin{equation}  
{\bf m}=\chi {\bf h} + \delta {\bf M}= \chi \left [ {\bf h} + \frac{\tau}{2}
  (\nabla \times {\bf v}) \times {\bf H}_0 \right ] . 
\label{konstit_s_lin}
\end{equation} 
The potentials $\psi$ and $\psi^{(2)}$ are solved by 
\begin{align} 
\psi&= -\frac{\tau \chi}{2 (1+\chi)} H_0 w + {\bar \psi}\,  
e^{kz} \, e^{i(kx-\omega t)}   
 \label{psi_s}\\ 
\psi^{(2)}&={\bar \psi^{(2)}} e^{-kz} e^{i(kx-\omega t)}. 
\end{align} 
The boundary conditions (\ref{b_bound},\ref{h_bound}) yield the as yet unknown  
pre-factors  
\begin{align} 
{\bar \psi}&=- \left [ \frac{\chi}{2+\chi} M_0 - 
 \frac{\tau \chi}{2+\chi} H_0 \left [ \nu k^2 (1-\frac{{\tilde q}}{k}) -  
i \omega \right ] \right ] {\bar \zeta}   
\label{koeff1_s}\\
{\bar \psi}^{(2)}&=\left [ \frac{1+\chi}{2+\chi}M_0 + 
 \frac{\tau \chi}{2+\chi} H_0 \nu k^2 (1-\frac{{\tilde q}}{k}) \right ] 
 {\bar \zeta}, 
\label{koeff2_s} 
\end{align} 
where ${\tilde q}=\sqrt{k^2-i \omega /[\nu(1+\kappa/4)]}$.  
>From Eq.~(\ref{psi_s}) and the expressions  
for the coefficients (\ref{koeff1_s},\ref{koeff2_s}) 
 it can be seen that the solution of the magnetic field 
equations couple to the velocity field $w$. This coupling is removed in
the absence of the magneto-dissipation  
($\tau=0$) and the result of the  
standard inviscid calculation  
\cite{rosensweig85} is recovered. These considerations complete the solution of the  
magnetic field problem in terms of the hydrodynamic variables.  
 
We now turn to the solution of the hydrodynamic problem.  
>From the linearized equations (\ref{ns_s},\ref{ns_l}) we eliminate the  
gradient terms by operating  twice with curl on the Navier-Stokes-equation.  
Projecting on the $z$-axis and using again Eq.~(\ref{konstit_s_lin}) we find 
\begin{equation} 
\left [ \rho \partial_t - (1+\frac{\kappa}{4}) \eta \nabla^2 \right ]  
\nabla^2 w= 0.  
\label{lin_ns_s}
\end{equation} 
Introducing again the plane wave ansatz for the lateral 
dependence of $w$, the vertical 
dependence is determined by 
\begin{equation} 
(\partial_z^2-k^2)(\partial_z^2-{\tilde q}^2) {\bar w}=0 
\label{eq_s}
\end{equation} 
with  
\begin{equation} 
{\tilde q}^2=k^2-\frac{i \omega}{\nu (1+\kappa/4)}. 
\end{equation} 
The proper solution of (\ref{eq_s}) remaining bounded  
for $z\to-\infty$ is  
\begin{equation} 
\wb(z)=A e^{kz}+B e^{{\tilde q}z} 
\end{equation}  
with the so far undetermined coefficients $A$ and $B$. It is always understood  
that the real part of ${\tilde q}$ is positive. 
Using the balance equation for normal stresses (\ref{normal_lin}) and the kinematic surface  
condition (\ref{kinemat_lin}) allows for to express $A$ and $B$ in terms of ${\bar \zeta}$,
resulting finally in Eq.~(\ref{D_s}).

\subsection{The hydrodynamic approach}
Again (\ref{h_bulk}) allows to introduce scalar magnetic potentials
\begin{eqnarray} 
\label{defpsi} 
\Ha^R+\Ha^D&=-\na \psi\\ 
\Ha^{(2)}&=-\na \psi^{(2)}\quad, 
\end{eqnarray} 
The linearization of Eq. (\ref{HDequ}) is in the present case
\begin{equation}
\Ha^D=\frac{\alpha}{\mu_0}\partial_t \B-\alpha\He\partial_z\ve\quad.
\end{equation} 
>From this equation and (\ref{b_bulk}) we find that due to the
incompressibility of the flow the magnetic potentials must be
harmonic. Therefore 
\begin{align} 
\psi&= {\bar \psi} \, e^{kz} \, e^{i(kx-\omega t)}\\
\psi^{(2)}&={\bar \psi^{(2)}} e^{-kz} e^{i(kx-\omega t)}. 
\end{align} 
Using the boundary conditions (\ref{b_bound},\ref{h_bound}) we find 
\begin{align} 
\psib&=\frac{\frac{\al}{k}\m\He\pa_z\wb(0) -(1-i\m\al\omega)\zetab M_0} 
                 {1+\m-i\m\al\omega} \\
\label{koeff1_l} 
\psib^{(2)}&=\frac{\frac{\al}{k}\m\He\pa_z\wb(0)+\m\zetab M_0} 
                 {1+\m-i\m\al\omega}\quad, 
\end{align} 
where $\pa_z\wb(0)$ denotes the derivative of $\wb(z)$ at $z=0$ and
$\mu_r=1+\chi$. Again it can be seen that the solution of the magnetic field 
equations couples to the velocity field $w$. In the absence of the
magneto-dissipation   
($\alpha=0$) this coupling is removed and the result of the  
standard inviscid calculation \cite{rosensweig85} is recovered. Moreover,
considering the low frequency approximation $\alpha\omega\ll 1$ one finds that 
the {\it reactive} magnetic field $\Ha^R$ is exactly the same as in the absence of field
dissipation. 
Eliminating from the linearized equations (\ref{ns_s},\ref{ns_l}) the 
gradient terms by operating twice with curl on the Navier-Stokes-equation and
projecting on the $z$-axis we find 
\begin{equation} 
(\rho \partial_t -  \eta\nabla^2) \nabla^2 w=  
- \alpha \He \partial_t \partial_z \nabla^2 b_z + \alpha \mu_0 H_{ext}^2 \partial_z^2 
\nabla^2 w  
\label{lin_ns_l} 
\end{equation} 
>From the fact that $\psi$ is harmonic, $\nabla^2 b_z$ can be replaced by  
\begin{equation} 
\nabla^2 b_z=\mu_0\m \frac{\al\He}{1-i\m\al\omega}\pa_z\nabla^2 w. 
\end{equation} 
Using again the plane wave ansatz for the lateral dependence of $w$, the
$z$-dependence is determined by  
\begin{equation} 
(\partial_z^2-k^2)(\partial_z^2-{\hat   q}^2) {\bar w}=0\quad, 
\label{eq_l} 
\end{equation} 
with  
\begin{equation} 
{\hat q}^2=\frac{k^2-\frac{i\omega}{\nu}}{1+\kappa}
\end{equation} 
with the real part of ${\hat q}$ positive. 
The proper solution for $\wb(z)$ remaining bounded for $z\to-\infty$ is hence
\begin{equation} 
\label{resw} 
\wb(z)=A e^{kz}+B e^{\hat{q}z} 
\end{equation}  
with the so far undetermined coefficients $A$ and $B$. Using now
$\alpha\omega\ll 1$ and expressing $\zetab$
in terms of $A$ and $B$ by using the kinematic boundary condition the 
requirement of  non-trivial solutions in $A$ and $B$ for the two stress
boundary conditions results finally in (\ref{D_l}).

\begin{figure} 
\psfig{file=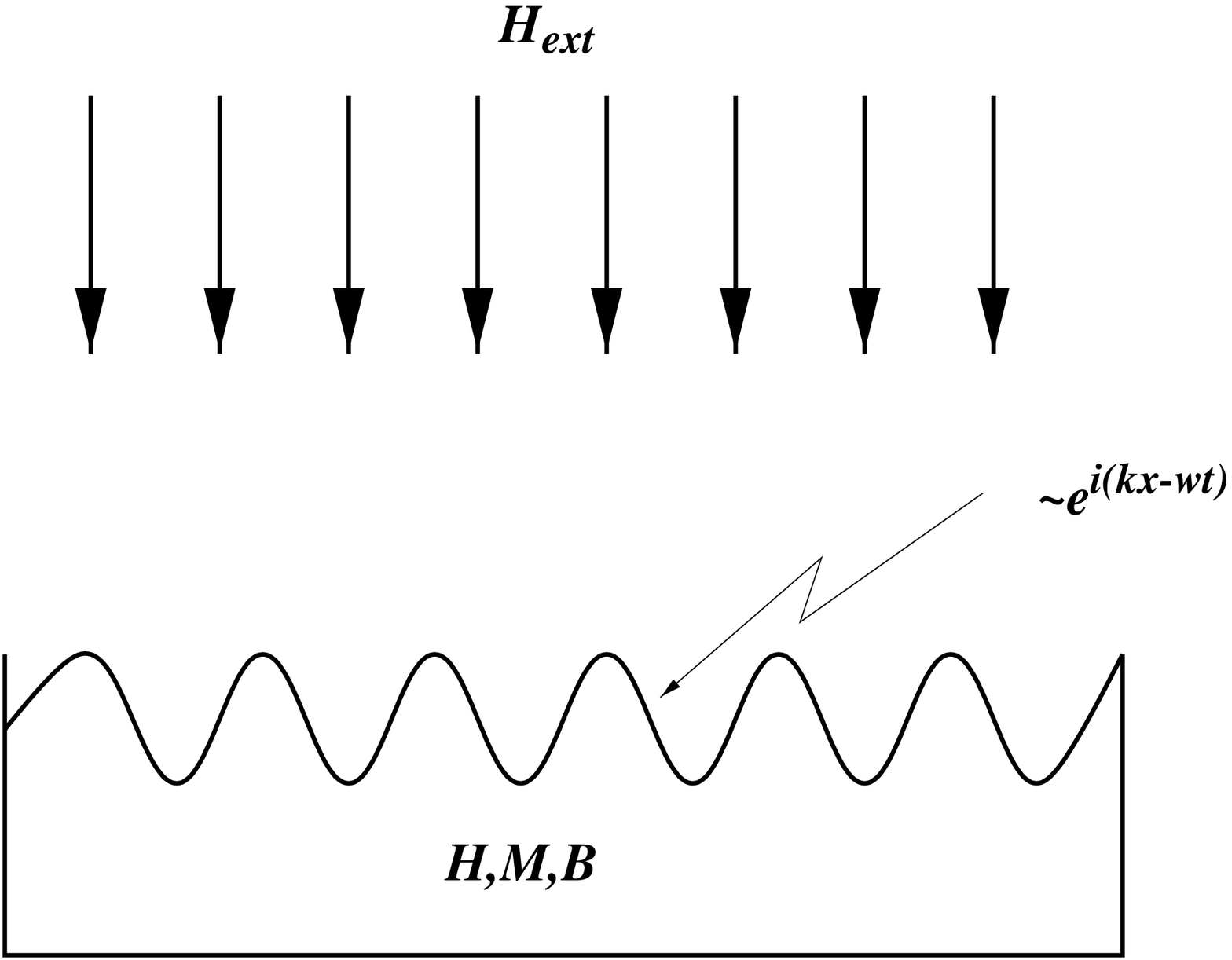,width=0.95\columnwidth}   
 \parbox[t]{8 cm}  
\caption[]   
{Sketch of the experimental setup. A ferrofluid of infinite lateral extension and 
infinite depth is exposed to a stationary magnetic field ${\bf H}_{ext}$ directed 
normally to the surface.}    
\label{fig1}   
\end{figure}  
\begin{figure}  
 \psfig{file=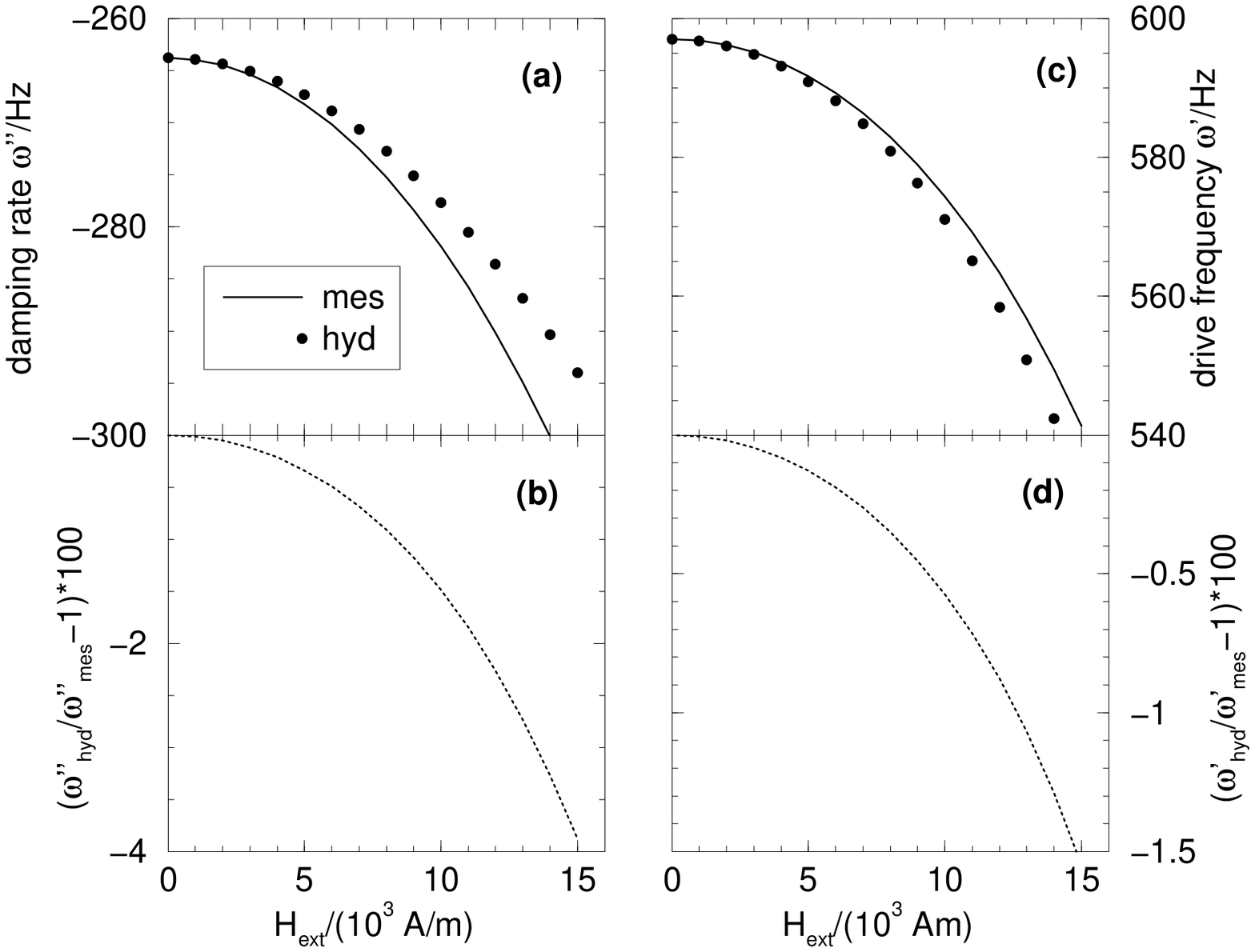,width=0.95\columnwidth}   
 \parbox[t]{8 cm}  
\caption[]   
{(a) The decay rate $\omega''(k,H_{ext})$ for free waves on the surface of a
ferrofluid exposed to a normal  magnetic ${\bf H}_{ext}$.  Solid lines (dots) denote 
the mesoscopic (hydrodynamic) theory. 
For the maximum investigated field intensity $H_{ext}= 
15 kA/m$ the relative deviation (b) does not exceed $4\%$. 
Parts (c,d) demonstrate how the wave excitation frequency $\omega'$ must be  
adapted to keep the the inviscid contribution $\omega_0(k,H_{ext})$ constant 
during the  
variation of the field intensity. For the evaluation of the curves we used 
Eqs.(\ref{XM_s},\ref{XM_l}) and the material
specifications for the ferrofluid APG J12  with an 
estimated
magnetic relaxation time of $\tau=3.8\times 10^{-5}s$.
}   
\label{fig2} 
\end{figure}   
  

\begin{thebibliography}{99}
\bibitem{footnote1} 
Note that the reactive contribution ${\bf M}^R=\chi {\bf H}^R$ of the 
hydrodynamic theory differs from the equilibrium magnetization 
${\bf M}_{eq}=\chi {\bf H}$ defined in the framework of the mesoscopic 
approach.
\bibitem{footnote3}The specifications for the ferrofluid APG J12 are as follows:  
Saturation magnetization $\mu_0 M_s=20mT$, viscosity $\eta=40mPas$, 
density $\rho=1.11g/cm^3$, volume fraction of magnetic material $\Phi_m=3.3\%$,  
susceptibility $\chi=0.8$, and surface tension $\gamma=0.034 N/m$.
\end{thebibliography}
\end{document}